# CHANNEL OPTIMIZED DISTRIBUTED MULTIPLE DESCRIPTION CODING

*Mehrdad Valipour, Farshad Lahouti*

valipour@ut.ac.ir, lahouti@ut.ac.ir
Wireless Multimedia Communications Laboratory
School of Electrical & Computer Eng.
College of Engineering
University of Tehran
Tel.: +98-21-82084314

*Abstract*

In this paper, channel optimized distributed multiple description vector quantization (CDMD) schemes are presented for distributed source coding in symmetric and asymmetric settings. The CDMD encoder is designed using a deterministic annealing approach over noisy channels with packet loss. A minimum mean squared error asymmetric CDMD decoder is proposed for effective reconstruction of a source, utilizing the side information (SI) and its corresponding received descriptions. The proposed iterative symmetric CDMD decoder jointly reconstructs the symbols of multiple correlated sources. Two types of symmetric CDMD decoders, namely the estimated-SI and the soft-SI decoders, are presented which respectively exploit the reconstructed symbols and a posteriori probabilities of other sources as SI in iterations. In a multiple source CDMD setting, for reconstruction of a source, three methods are proposed to select another source as its SI during the decoding. The methods operate based on minimum physical distance (in a wireless sensor network setting), maximum mutual information and minimum end-to-end distortion. The performance of the proposed systems and algorithms are evaluated and compared in detail.

*EDICS*: SEN-DCSC Distributed channel and source coding

# I. INTRODUCTION

The robustness of multiple descriptions (MD) coding against packet loss (PL) and its multi-rate capability makes it an attractive solution for multimedia compression and communications over lossy networks [1]. Noting, in addition, the simplicity of MD, it also suits wireless sensor network applications. An MD encoder generates several descriptions of a source symbol, and depending on the set of such descriptions available at the decoder, the source symbol is reconstructed at the decoder with different qualities. Distributed source coding (DSC) allows for exploiting the dependency of distributed sources for efficient compression and/or transferring the source codec complexity from the encoder to the decoder. This is of particular interest, e.g., in distributed monoview and multiview video coding for wireless up-link applications (see, e.g., [2] and the references therein) and of course wireless sensor networks. The communications over wireless packet networks, or combination of wireline and wireless networks is subject to noise and PL. Therefore, the design of a distributed multiple description system could play an enabling role in such prospective systems. A distributed multiple description system refers to a system with multiple descriptions as source encoders for several correlated and physically distributed sources and a source decoder for efficient reconstruction of all sources. This paper deals with the design, performance evaluation and analysis of distributed multiple description systems in presence of noise and PL.

According to the Slepian-Wolf theorem, separate lossless compression of two discrete correlated sources, could be as optimal as their joint encoding [3]. For jointly Gaussian sources, the Wyner-Ziv theorem states that, lossy compression of a source when another correlated source is available only at the decoder could be as optimal as it were also available at the encoder [4]. In this scenario that is referred to as the asymmetric DSC setting, the correlated source at the decoder serves as side information (SI) for decoding. A general scenario, known as multi-terminal source coding, is introduced in [5], where a number of sources transmit their correlated data to a number of destinations. Several inner and outer rate-distortion bounds are obtained for special cases in multi-terminal source coding, although the problem is still widely open. The interested reader is referred to [6-10] and the references therein for more details.

Recently much effort has been devoted to designing source encoders for multi-terminal source coding over noiseless channels, e.g., [8][11-13] and the references therein. In [14], two scalar quantizers are designed for two distributed correlated sources (symmetric setting) aiming at minimizing the overall reconstruction distortion at a single destination over noisy channels without PL. A distributed multi-stage coding of correlated sources is studied in [15].

A multiple description scalar quantizer composed of a scalar quantizer followed by an index assignment (IA) table is designed in [16] for robust communications over channels with packet loss. The design of a MD vector quantizer using deterministic annealing [17] is presented in [18], and based on simulated annealing in [19]. Recently, the design of multiple descriptions over noisy channels with and without PL has received noticeable attention, e.g., [20-21].

In a scheme suggested in [22], in addition to two descriptions of a source sample, extra information generated by a WZ encoder, is transmitted for enhanced reconstruction in presence of packet loss. The authors consider the design of multiple description vector quantization (MDVQ) with SI available at the decoder in [23]. In [24], a MD with SI is considered using a non-optimized MDVQ and parity-based turbo codes as the Slepian-Wolf encoders over noiseless channels.

To the best of our knowledge, this work is the first on the design of channel optimized distributed multiple description vector quantization (CDMD). For efficient distributed source encoding and joint reconstruction of correlated sources over noisy channels with packet loss, CDMD source encoders and an iterative symmetric CDMD decoder are designed. To this end, first an asymmetric CDMD (ACDMD) scheme is presented. In this case, the encoded descriptions are reconstructed over noisy channels with packet loss at a single MMSE decoder exploiting SI. Next based on the results, two symmetric iterative CDMD decoders, namely the estimated-SI and the soft-SI decoders, are proposed in which for the reconstruction of each source at the decoder one of the other sources are assumed as SI. The two schemes respectively consider SI as either the computed reconstructed signal or the a posteriori probability produced by the ACDMD decoder of the corresponding source in each iteration. Three schemes of minimum physical distance, maximum mutual information and minimum distortion are studied for selecting the proper source as side information.

Simulation results of the asymmetric CDMD system indicate a noticeable gain in comparison to classic MDVQ or channel optimized MDVQ, while being only 2 dB away from the rate-distortion bound for a two-description system with side information. The effects of side information quantization rate, bit error rate and packet loss probability on the performance are studied. Also, it is shown that underestimating the correlation between the sources for CDMD encoder design is less disturbing than its overestimation in terms of the overall performance. The performance of the proposed symmetric CDMD decoders (estimated-SI and soft-SI) is thoroughly investigated for randomly distributed correlated sources in presence of noise and PL considering both cases of transmission over binary symmetric channels (BSC) and additive white Gaussian noise (AWGN) channels. The effectiveness of the proposed symmetric CDMD decoder in large sensor networks as a function of number and density of the nodes is also studied.

This paper is organized as follows. The system model and the problem formulation are presented in section II. In section III, the asymmetric CDMD is introduced and the design of ACDMD source encoder is considered. The symmetric CDMD is elaborated in section IV. Side information selection strategies are presented and studied in Section V. Finally in section VI, the designed CDMD is evaluated.

## II. SYSTEM MODEL

Fig. 1 demonstrates the block diagram of a symmetric CDMD. Symbols generated by $N^u$ memoryless distributed correlated sources are encoded independently using CDMD source encoders. The indices at the output of the source encoders are transmitted to destination node(s) over a noisy network with packet loss. An iterative CDMD decoder jointly reconstructs the symbols of the sources. This decoder consists of $N^u$ asymmetric CDMD decoders and a side information controller. Each asymmetric decoder reconstructs the symbols of a specific source; and the side information controller selects another source as its SI, and sets the dependency considered between the source and SI symbols at the ACDMD iterations.

In order to design the symmetric CDMD decoder, first an ACDMD system is considered when there is only one source and the SI is available at the decoder. A block diagram for the ACDMD is illustrated in Fig. 2. At the encoder, similar to an MDVQ, first a source symbol is quantized; then each cell of the quantizer is mapped to a set of descriptions (indices) using an index assignment table. Note that one or more cells of the quantizer could be assigned the same index, similar to a Wyner-Ziv quantizer. The generated descriptions indices are transmitted to the receiver over independent channels that are subject to noise and packet loss [25].

Then using the noisy received descriptions and the SI available at the decoder, an asymmetric MMSE CDMD decoder reconstructs the source samples.

In the following, the problem is formulated and the notations are defined.

*A. Problem Formulation*

In this paper, capital letters, e.g., $X$, represent random variables, while small letters, e.g., $x$, represent a realization. We replace the probability $P(X = x)$ by $P(X)$ in most instances, when it does not lead to confusion. A bold faced letter shows a vector. Letters $u$ and $m$ as scripts indicate the index of source and the index of description, respectively.

Suppose $\boldsymbol{X}^u$, $u \in 1, \cdots, N$ be the $u^{th}$ source symbol, where $N$ is the number of distributed sources. The $\boldsymbol{X}^u$ is mapped to one of $K^u$ cells of the $u^{th}$ vector quantizer corresponding to the codeword $\widetilde{\boldsymbol{X}}_k^u, k \in \{1, \ldots, K^u\}$. This may be described by the probability distribution function $P(\widetilde{\boldsymbol{X}}_k^u|\boldsymbol{X}^u)$ for each $k$, where,

$$P(\widetilde{\boldsymbol{X}}_k^u|\boldsymbol{X}^u) = \begin{cases} 1 & \text{if } \boldsymbol{X}^u \in \mathcal{V}^u(k) \\ 0 & \text{Otherwise.} \end{cases} \qquad (1)$$

The $\mathcal{V}^u(k)$ and $P^u(k)$ indicate the $k^{th}$ quantizer cell (Voronoi region) of the $u^{th}$ source and its corresponding probability, i.e., $P^u(k) = \int_{x^u \in \mathcal{V}^u(k)} P(x^u) d\,x^u$. The mapping from quantizer cells to descriptions indices is determined by an index assignment table described by the probability distribution function $P(\boldsymbol{I}^u|\widetilde{\boldsymbol{X}}_k^u)$ for each $\boldsymbol{I}^u = (I_1^u, \cdots, I_{M^u}^u)$ and $k$, where $M^u$ is the number of descriptions of the source and $\sum_{all\,I^u} P(\boldsymbol{I}^u|\widetilde{\boldsymbol{X}}_k^u) = 1$ for each $k$ and $u$. The $I_m^u$ is the index of the $m^{th}$ description and is selected from a set with $N_m^u$ elements. This index may be represented by $\lceil \log_2 N_m^u \rceil$ bits. While $P(\boldsymbol{I}^u|\widetilde{\boldsymbol{X}}_k^u) \in \{0,1\}$ during the operations, as we shall elaborate in the sequel, we consider it as $0 \leq P(\boldsymbol{I}^u|\widetilde{\boldsymbol{X}}_k^u) \leq 1$ during the design.

The descriptions indices of the $u^{th}$ source are converted to binary sequences and transmitted over $M^u$ equivalent independent and memoryless noisy channels with packet loss [25].The received sequence at the input of the $u^{th}$ ACDMD decoder is denoted by $\boldsymbol{J}^u = (J_1^u, \cdots, J_{M^u}^u)$, where $J_m^u$ is the channel output corresponding to the $m^{th}$ description of the $u^{th}$ source. The binary sequence $\boldsymbol{Q}^u = \{Q_1^u, \cdots, Q_{M^u}^u\}$, indicates the status of descriptions at the decoder, i.e., $Q_m^u = 0$ indicates that the $m^{th}$ description of the $u^{th}$ source has been subject to packet loss and is unavailable at the decoder. The probability $P(\boldsymbol{Q}^u = \boldsymbol{q}^u)$ or simply $P(\boldsymbol{Q}^u)$ indicates the channels packet loss probability. The channels of different descriptions are assumed independent, therefore,

$$P(\boldsymbol{Q}^u = \boldsymbol{q}^u) = \prod_{m=1}^{M^u} P(Q_m^u = q_m^u), \tag{2}$$

$$P(\boldsymbol{J}^u|\boldsymbol{I}^u, \boldsymbol{Q}^u) = \prod_{m=1}^{M^u} P(J_m^u|I_m^u, Q_m^u), \tag{3}$$

where,

$$P(Q_m^u = q_m^u) = (1 - q_m^u)\mu_m^u + q_m^u(1 - \mu_m^u), \tag{4}$$

and $\mu_m^u$ is the packet loss probability of the channel of the $m^{th}$ description for the $u^{th}$ source. For BSC with the bit error rate equal to $P_m^u$, we have:

$$P(J_m^u|I_m^u, Q_m^u) = \begin{cases} (P_m^u)^{d_H(I_m^u, J_m^u)}(1 - P_m^u)^{\lceil \log_2 N_m^u \rceil - d_H(I_m^u, J_m^u)} & Q_m^u = 1 \\ 1/N_m^u & Q_m^u = 0 \end{cases}, \tag{5}$$

where $d_H(\boldsymbol{I}, \boldsymbol{J})$ is the hamming distance between $\boldsymbol{I}$ and $\boldsymbol{J}$; and for an AWGN channel with the noise power spectral density of $N_0/2$ and BPSK transmission modulation, $P(J_m^u|I_m^u, Q_m^u)$ is equal to:

$$P(J_m^u|I_m^u, Q_m^u) = \begin{cases} 1/\sqrt{\pi N_0} \exp(-4\, d_H^2(I_m^u, J_m^u)/N_0) & Q_m^u = 1 \\ 1/N_m^u & Q_m^u = 0 \end{cases}, \tag{6}$$

Let $\boldsymbol{Y}^u$ and $\widehat{\boldsymbol{X}}^u(\boldsymbol{J}^u|\boldsymbol{Y}^u, \boldsymbol{Q}^u)$ denote the side information available at $u^{th}$ ACDMD decoder, and the corresponding decoded symbol, respectively. The average distortion is then defined as,

$$D^{av} \stackrel{\text{def}}{=} \frac{1}{N} \sum_{u=1}^{N} d(\widehat{\boldsymbol{X}}^u(\boldsymbol{J}^u|\boldsymbol{Y}^u, \boldsymbol{Q}^u), \boldsymbol{X}^u). \tag{7}$$

When a mean squared error (MSE) distortion criterion is used, the distortion is written as:

$$D^{av} = \frac{1}{N} \sum_{u=1}^{N} \left(\boldsymbol{X}^u - \widehat{\boldsymbol{X}}^u(\boldsymbol{J}^u|\boldsymbol{Y}^u, \boldsymbol{Q}^u)\right)^2. \tag{8}$$

The main goal of this paper is to design the CDMD encoder and decoder for robust communications over noisy channels with packet loss such that the average reconstruction (MSE) distortion is minimized. In the sequel, we will drop the superscript $u$ (the source index) when it does not cause an ambiguity.

### III. ASYMMETRIC CDMD

In this section, an asymmetric CDMD is presented for robust communications in presence of noise and packet loss. To this end, first the average distortion is described for a MD system in this setting. Next the source decoder of the proposed channel optimized distributed M-description vector quantizer is introduced, and finally a procedure for designing CDMD encoder based on a deterministic annealing algorithm is presented. For simplicity, the superscript $u$ is omitted in this part.

*A. Evaluating Distortion*

When the mean squared error is the distortion criterion, the average distortion $D^{av}$ is decomposed into two parts: channel distortion $D^{Ch}$ due to noise and packet loss, and source encoder distortion $D^{SE}$.

*Proposition 1*: The average MSE distortion in an asymmetric CDMD decoder, when the conditional probability distribution of source symbol given SI is $f(X|Y)$, may be described as

$$D^{av} = D^{SE} + D^{Ch}, \qquad (9)$$

where,

$$D^{SE} = \int f(Y) \left( \sum_I \int f(X,I|Y)(X - C(I|Y))^2 dX \right) dY, \qquad (10)$$

$$D^{Ch} = \sum_Q P(Q) D^{Ch}_Q, \qquad (11)$$

in which,

$$D^{Ch}_Q = \sum_J \sum_I P(J|Q,I) \int f(Y) P(I|Y) \left( C(I|Y) - \widehat{X}(J|Y,Q) \right)^2 dY. \qquad (12)$$

$$C(I|Y) \stackrel{\text{def}}{=} \int X f(X|Y,I) dX, \qquad (13)$$

$$P(I|Y) = \sum_k P(I|\widetilde{X}_k) P(\widetilde{X}_k|Y), \qquad (14)$$

where $C(I|Y)$ and $P(I|Y)$ (both stored at the decoder) are the decoder codebook and the dependency probabilities of the ACDMD source codec given side information; and,

$$P(\widetilde{X}_k|Y) = \int P(\widetilde{X}_k|X) f(X|Y) dX, \qquad (15)$$

$$f(X|Y,I) = \frac{f(X|Y)}{P(I|Y)} \left[ \sum_k P(I|\widetilde{X}_k) P(\widetilde{X}_k|X) \right]. \qquad (16)$$

*Proof:* The proof is presented in Appendix A. ▫

As evident in (10)-(12), the $D^{SE}$ only depends on the CDMD source encoder and joint probability distribution of source and side information. Also, the channel distortion $D^{Ch}$ indicates the effect of channel disturbances reflected by the channel transition probability, $P(J|Q = q, I)$. Note that $D^{Ch}$ is still influenced by the source encoder characteristics.

*B. Reconstruction of Source Symbols*

As mentioned, the MMSE ACDMD decoder reconstructs the source symbols based on the noisy output of the lossy channels exploiting the side information available at the decoder. The following proposition presents the decoder.

*Proposition 2*: In an asymmetric CDMD decoder, the reconstructed signal $\hat{X}(J|Y, Q)$ based on a MMSE criterion is given by,

$$\hat{X}(J|Y, Q) = \sum_{I} P(I|Y, Q, J) C(I|Y), \qquad (17)$$

where,

$$P(I|Y, Q, J) = \frac{P(J|Q, I) P(I|Y)}{P(J|Y, Q)}, \qquad (18)$$

$$P(J|Y, Q) = \sum_{I} P(J|Q, I) P(I|Y), \qquad (19)$$

and $P(I|Y)$ and $C(I|Y)$ are introduced in (13) and (14).

*Proof*: The proof is presented in Appendix B. ▫

As evident in (13)-(19), the reconstructed signal is a function of the quantizer $P(\widetilde{X}_k|X)$, the index assignment $P(I|\widetilde{X}_k)$, the channel transition probability $P(J|Q, I)$, and the joint probability distribution of $X$ and $Y$, $f(X, Y)$.

*Remark*- In general, the side information, being a continuous value, is quantized to $N^{SI}$ levels. The effect of the SI quantizer size is investigated in section VI. The MMSE decoder using quantized SI, $\widetilde{Y}_k$, is similar to that presented in Proposition 2 with continuous SI, $Y$. The difference is that $f(X|Y)$ in (15)-(16) is replaced by,

$$f(X|\widetilde{Y}_k) = f(X) \int_{Y \in V_y(k)} f(Y|X) \, dY \Big/ P(\widetilde{Y}_k), \qquad (20)$$

in which $V_y(k)$ indicates the Voronoi region of the SI quantizer.

### C. Source Encoder Design

The SE design objective in the ACDMD setting is to find the IA table, represented by $P(I|\widetilde{X}_k)$, such that $D_{av}$ in Proposition 1 is minimized. During the SE design, it is assumed that the quantizer, characterized by $P(\widetilde{X}_k|X)$, and the joint probability distribution of source and side information, $f(X, Y)$, are fixed.

The deterministic annealing [17] is an optimization technique, which is initialized probabilistically, and then, in each iteration, the level of randomness is gradually decreased until a non random solution is obtained. Specifically, for ACDMD design, the quantizer cells are assigned to the cells of the IA table based on the probabilities $P(I|\widetilde{X}_k)$ for each $I$ and $k$. In each step, an IA is designed for minimizing the distortion given a specific level of randomness for the IA table. This level of randomness is quantified by $\mathcal{H}(I|K)$, where,

$$\mathcal{H}(I|K) \triangleq -\sum_I \sum_k P(I, \widetilde{X}_k) \log P(I|\widetilde{X}_k). \qquad (21)$$

The value of $\mathcal{H}(I|K)$ is reduced from the value corresponding to a completely random IA to zero for a non-random IA. The IA table, obtained in one step, is used for the initialization of the next step. As a result, when the design algorithm terminates, $P(I|\widetilde{X}_k)$ for each $I$ and $k$ takes a value of either zero or one. Therefore, the goal of the intermediate optimization step is as follows,

$$P(I|\widetilde{X}_k) = \arg\min D^{av}. \qquad (22)$$
$$\text{s.t.} \begin{cases} \sum_I P(I|\widetilde{X}_k) = 1, \forall k \in \{1, \dots, K\} \\ \mathcal{H}(I|K) = H_0 \end{cases}$$

The Lagrangian based on (22) is written as,

$$\mathcal{L}(T, \lambda_1, \dots, \lambda_K) = D_{av} - T\mathcal{H}(I|K) + \sum_k \left[\sum_I P(I|\widetilde{X}_k)\right] \lambda_k, \qquad (23)$$

Where $T$ and $\lambda_1, \dots, \lambda_K$ are the Lagrangian multipliers. We next set the derivative of (23), with respect to $P(I|\widetilde{X}_k)$, equal to zero and consider the first $K$ constraints in (22) to obtain

$$P(I|\widetilde{X}_k) = \frac{\exp\left(-\mathcal{W}(I,k)/(T.P(\widetilde{X}_k))\right)}{\sum_{I'} \exp\left(-\mathcal{W}(I',k)/(T.P(\widetilde{X}_k))\right)}, \quad (24)$$

where,

$$\mathcal{W}(I,k) \stackrel{\text{def}}{=} \frac{\partial D_{av}}{\partial P(I|\widetilde{X}_k)} \quad (25)$$

$$= \sum_{Q \in \{0,1\}^M} P(Q) \sum_J P(J|Q,I) \int P(\widetilde{X}_k|X) f(X) \int f(Y|X)(X - \widehat{X}(J|Y,Q))^2 dY \, dX.$$

Note that one could use (24) and the last constraint in (22) to obtain $P(I|\widetilde{X}_k)$ in terms of $\mathcal{H}(I|K)$ or $H_0$. Alternatively, we can use (24) and take $T$ as the measure of randomness. The CDMD design procedure, with deterministic annealing algorithm, may now be summarized as follows. First a vector quantizer $P(\widetilde{X}_k|X)$ is selected, and the IA table $P(I|\widetilde{X}_k)$ is randomly initialized. Starting with a sufficiently high $T$, the IA table is updated using (24)-(25), and the distortion is computed using (9)-(12). Note that $P(I|\widetilde{X}_k)$ obtained in the previous step is used for computing $\widehat{X}(J|Y,Q)$. If the obtained average distortion improves less than a certain threshold, or the number of iterations exceeds a specific limit, $T$ is reduced, otherwise the IA table is updated again. Whenever $T$ is smaller than a certain small threshold, the algorithm is terminated.

## IV. SYMMETRIC CDMD

As described in section II, in a symmetric CDMD, the source symbols, generated by $N^u$ correlated sources are encoded separately and reconstructed jointly by an iterative symmetric CDMD decoder. In this part, two schemes for the iterative symmetric CDMD decoding are proposed. The resulting decoders are referred to as the "estimated-SI" decoder and "soft-SI" decoder, and are distinct in the way they exploit the side information in iterations.

### A. Symmetric CDMD Decoder using Estimated-SI

Based on the symmetric CDMD system model described in section II, suppose the $s^{th}$ source is selected as the side information source of the $u^{th}$ source during the decoding. The SI source selection strategies are studied in the section V. The $u^{th}$ ACDMD decoder reconstructs the $u^{th}$ source symbols using Proposition 2 considering an MMSE criterion and the side information. In the next iteration, as shown in Fig. 1, the reconstructed values of the $s^{th}$ ACDMD decoder are quantized and used as the side information of the $u^{th}$ ACDMD decoder.

As the ACDMD decoders only provide reconstructed values following the first iteration, no estimated symbols (producing SI, $Y$) is available at the ACDMD decoders in the first iteration. Therefore, for the first iteration $f(X|Y)$ is replaced by $f(X)$ for computing $P(I|Y)$ and $C(I|Y)$ in Proposition 2. For the subsequent iterations, given the estimated symbols of the source, $f(X|Y)$ is used at the ACDMD decoder.

## B. Symmetric CDMD Decoder using Soft-SI

According to Fig. 1, if the $s^{th}$ source is selected as the side information source of the $u^{th}$ source, the soft-SI decoder is an iterative symmetric CDMD decoder in which the a posteriori probabilities produced by the $s^{th}$ ACDMD decoder, i.e., the set of $Y^u = \{P(I^s|Y^s, Q^s, J^s), \forall I^s\}$, are used as the side information of the $u^{th}$ ACDMD decoder. In the soft-SI decoder, computing the reconstructed values is not necessary until the last iteration. In the next two Propositions, a posteriori probabilities $P(I^u|Y^u, Q^u, J^u)$, and reconstructed signals $\hat{X}^u(J^u|Y^u, Q^u)$ are computed in this setting.

As there are no a posteriori probabilities available in the first iteration, $P(I^s|Y^s, Q^s, J^s)$ is considered uniformly distributed first; and for the next iterations, the ACDMD decoder operates with the exact dependency between source and SI.

*Proposition 3*: For the symmetric CDMD decoder using soft-SI, the a posteriori probabilities are computed as,

$$P(I^u|Y^u, Q^u, J^u) = \frac{P(J^u|I^u, Q^u)P(I^u|Y^u)}{P(J^u|Y^u, Q^u)}, \quad (26)$$

where,

$$P(J^u|Y^u, Q^u) = \sum_{I^u} P(J^u|I^u, Q^u)P(I^u|Y^u), \quad (27)$$

$$P(I^u|Y^u) = \sum_{I^s} P(I^s|Y^s, Q^s, J^s) \sum_{k} P(I^u|\tilde{X}_k^s)P(\tilde{X}_k^s|I^s), \quad (28)$$

in which,

$$P(I^u|\tilde{X}_k^s) = \sum_{l} P(I^u|\tilde{X}_l^u)P(\tilde{X}_l^u|\tilde{X}_k^s), \quad (29)$$

and,

$$P(\widetilde{X}_l^u|\widetilde{X}_k^s) = \int_{X^u} f(\widetilde{X}_l^u|x^u) \int_{X^s} f(x^s|\widetilde{X}_k^s) f(x^u|x^s) dx^s\, dx^u. \tag{30}$$

*Proof*: The proof is presented in Appendix C.  □

*Proposition 4*: For the symmetric CDMD decoder using a posteriori probabilities $\{P(I^s|Y^s,Q^s,J^s), \forall I^s\}$ as soft-SI, the reconstructed signal $\widehat{X}^u(J^u|Y^u,Q^u)$ is given by,

$$\widehat{X}^u(J^u|Y^u,Q^u) = \sum_{all\, I^u} P(I^u|Y^u,Q^u,J^u) C(I^u|Y^u), \tag{31}$$

where $P(I^u|Y^u,Q^u,J^u)$ is computed using Proposition 3 and $C(I^u|Y^u)$ is obtained by,

$$C(I^u|Y^u) = \frac{1}{P(I^u|Y^u)} \sum_{I^s} P(I^s|Y^s,Q^s,J^s) \sum_{k} P(\widetilde{X}_k^s|I^s)\, P(I^u|\widetilde{X}_k^s) C(I^u|\widetilde{X}_k^s), \tag{32}$$

where $C(I^u|\widetilde{X}_k^s)$ is defined as,

$$C(I^u|\widetilde{X}_k^s) \stackrel{\text{def}}{=} \int_{X^u} x^u\, f(x^u|\widetilde{X}_k^s,I^u) dx^u, \tag{33}$$

and,

$$f(x^u|\widetilde{X}_k^s,I^u) = \frac{f(x^u|\widetilde{X}_k^s)}{P(I^u|\widetilde{X}_k^s)} \sum_l P(I^u|\widetilde{X}_l^u) P(\widetilde{X}_l^u|x^u). \tag{34}$$

*Proof*: The proof is presented in Appendix D.  □

## V. SELECTING THE SIDE INFORMATION SOURCE

As mentioned previously, in both schemes proposed for symmetric CDMD decoding, for each source, another source is selected as the side information source. Hence, the source symbols are reconstructed utilizing either the estimated value or the soft value of the ACDMD decoder corresponding to the SI source in an iterative manner.

We consider three criteria for selecting the side information source: minimum physical distance (in a sensor network scenario), maximum mutual information, and minimum distortion. In the following, these three methods are elaborated in detail.

*A. Minimum Physical Distance Criterion*

Consider a cluster of data gathering wireless sensor nodes, in which the nodes sense a natural phenomenon, e.g., temperature, and communicate it over lossy channels to a data gathering node or the cluster head. In such a setting, it is typical that the correlation (dependency) between the symbols of any two sources is larger when their physical distance is smaller. Thus, one criterion for selecting the side information source is that the $s^{th}$ source is selected as the SI source of the $u^{th}$ source if:

$$s = \arg\min_{t} \text{dist}(u, t), \tag{35}$$

where $\text{dist}(u, t)$ is the physical distance between the sources $u$ and $t$. Although selecting the SI source in this scheme is computationally simple once the distances are known, however, it does not take the reliability of the channels from sources to the data gathering node into account. Therefore, selecting the side information source with this criterion is not optimal. For example, source A may be close to source B but all the descriptions of source A are lost over the channel; using the minimum physical distance criterion, the source A is selected as the SI of source B, despite its lost descriptions. In the following, two other schemes are introduced for this purpose, which are more computationally complex, but provide better performance.

B. *Maximum Mutual Information Criterion*

To incorporate both the effect of dependency of data of different distributed sources, and their corresponding channel status, one may consider the mutual information between the received signal from a source and that due to any possible SI source as a criterion for SI source selection. In other words, considering Fig. 3, the $s^{th}$ source is selected as the SI source of the $u^{th}$ source if given the packet loss status of the channels,

$$s = \arg\max_{t} I(\boldsymbol{J}^u; \boldsymbol{J}^t | \boldsymbol{Q}^u = \boldsymbol{q}^u, \boldsymbol{Q}^t = \boldsymbol{q}^t), \tag{36}$$

where $I(\boldsymbol{X}; \boldsymbol{Y})$ is the mutual information between $\boldsymbol{X}$ and $\boldsymbol{Y}$. The mutual information in (36) is computed in Appendix E, and takes into account the source distribution, the quantizer and the IA table, as well as the PL status of channels and the statistics of the noisy channel.

C. *Minimum Distortion Criterion*

The proposed distributed MD decoders are designed to minimize the average (MSE) distortion. Therefore, here we consider a scheme for selecting the side information source such that the overall distortion is minimized. In this method, the $s^{th}$ source is selected as the SI source of the $u^{th}$ source when,

$$s = \arg\min_t D^{av}_{Y=(J^t,Q^t)} \quad , \tag{37}$$

where $D^{av}_{Y=(J^t,Q^t)}$ is the average distortion subject to packet loss status and the received signals of the SI, i.e.,

$$D^{av}_{Y=(J^t,Q^t)} = \mathrm{E}\left(\left(X^u - \widehat{X}^u(J^u|Y=(J^t,Q^t),Q^u)\right)^2\right). \tag{38}$$

In other words, as shown in Fig. 3, the noisy output, $J^t$, and the packet loss status, $Q^t$, of channels related to SI are directly used as the side information. The reconstructed value, $\widehat{X}^u(J^u|Y=(J^t,Q^t),Q^u)$, is computed using Proposition 5.

*Proposition 5*: The reconstructed signal of the MMSE ACDMD decoder of the $u^{th}$ source, $\widehat{X}^u(J^u|Y=(J^t,Q^t),Q^u)$, when the partial side information is available at the decoder $Y=(J^t,Q^t)$, shown in Fig. 3, is given by,

$$\widehat{X}^u(J^u|Y=(J^t,Q^t),Q^u) = \sum_{all I^u} P(I^u|J^u,J^t,Q^u,Q^t)C(I^u|J^t,Q^t), \tag{39}$$

where,

$$P(I^u|J^u,J^t,Q^u,Q^t) = \frac{P(J^u|I^u,Q^u)P(I^u|J^t,Q^t)}{P(J^u|J^t,Q^u,Q^t)}, \tag{40}$$

$$P(J^u|J^t,Q^u,Q^t) = \sum_{I^u} P(J^u|I^u,Q^u)P(I^u|J^t,Q^t), \tag{41}$$

$$P(I^u|J^t,Q^t) = \sum_l P(\widetilde{X}^u_l|J^t,Q^t)P(I^u|\widetilde{X}^u_l), \tag{42}$$

$$C(I^u|J^t,Q^t) = \frac{1}{P(I^u|J^t,Q^t)}\sum_k P(\widetilde{X}^t_k|J^t,Q^t)P(I^u|\widetilde{X}^t_k)C(I^u|\widetilde{X}^t_k), \tag{43}$$

$$P(\widetilde{X}^t_k|J^t,Q^t) = \frac{P(\widetilde{X}^t_k)}{P(J^t|Q^t)}\sum_{I^t} P(J^t|I^t,Q^t)P(I^t|\widetilde{X}^t_k), \tag{44}$$

$$P(J^t|Q^t) = \sum_k P(\widetilde{X}^t_k)\sum_{I^t} P(J^t|I^t,Q^t)P(I^t|\widetilde{X}^t_k), \tag{45}$$

and $P(I^u|\widetilde{X}^t_k)$ and $C(I^u|\widetilde{X}^t_k)$ are computed using (28) and (32), respectively.

*Proof:* The proof of (43) is similar to that of (32) in Appendix D. The proofs of (39)-(42) and (44)-(45) follow those of propositions 1 and 2 in Appendices A and B. ☐

For computing the average distortion in this method, the reconstructed values of the ACDMD decoder are required; therefore, the computational complexity in this case is more than the other two methods. As we shall see in the sequel, this added complexity leads to better performance. In section VI, the performance of the three proposed schemes for selecting the side information are investigated and compared.

## VI. CDMD Performance Evaluation

In this section, the performance of the proposed schemes for channel optimized distributed multiple description vector quantization is investigated over a network subject to noise and packet loss in both asymmetric and symmetric scenarios. The sensitivity of the performance to different systems and design parameters are also studied.

The reference setting for the experiments is as follows. A Lloyd scalar quantizer with 256 levels is used in a two description CDMD system. The number of indices for each description, $N_m$, $m = 1,2$, is 8. The typical value of packet loss rate is set to 0.05. The SI quantizer has 128 levels. The subscript *sim* indicates the ones obtained using simulations. The distortion is measured in dB. The sources are considered zero-mean Gaussian random variables with unit variance. For asymmetric CDMD, the correlation coefficient between the source and the SI is denoted by $\rho_{xy}^{real}$; and $\rho_{xy}^{enc}$ and $\rho_{xy}^{dec}$ refer to the correlation coefficients between $X$ and $Y$ used during the design of the encoder and decoder, respectively.

For performance evaluation of the symmetric CDMD, we consider a data gathering wireless sensor network setting, where the nodes are distributed randomly and uniformly in a unit square. The correlation coefficient between the sources $u$ and $s$, $\rho_{x_u x_s}^{real}$, with physical distance $d$ is set to $\exp(-d/\alpha)$; Therefore, the larger the distance between the sources the smaller their correlation. The parameter $\alpha$ is set to 2.

### A. Results for Asymmetric CDMD

In this section, first performance bound of the asymmetric MD with SI at the decoder is introduced. Then the performance of the proposed asymmetric CDMD is compared with the performance bound; and improvement of the end-to-end performance due to proper design of source encoder and decoder in the presence of side information is studied. Next the effects of bit error rate and packet loss probability of channels, and the SI quantization rate on the performance are evaluated via simulations. Also, the sensitivity of the average CDMD distortion to the mismatch of the assumed and exact values of dependancy (correlation) between the symbols of source and SI is investigated.

*1) Performance Bound for Asymmetric MD with SI and Noiseless Channels*

The information theory bound for multiple descriptions with side information is obtained only for a special case that is a two description system with SI over noiseless channels with PL. When the channels are noiseless, and the side information $Y$, is available at the decoder, the set of all achievable distortion region [26] for a given rate pair $(R_1, R_2)$ (bits) is given by the triple $(D_1, D_2, D_{12})$ satisfying [27],

$$D_1 \geq \beta . 2^{-2R_1}, \tag{46}$$

$$D_2 \geq \beta . 2^{-2R_2}, \tag{47}$$

$$D_{12} \geq \beta . 2^{-2(R_1+R_2)} / \left(1 - \left(\sqrt{\Pi} - \sqrt{\Delta}\right)^2\right), \tag{48}$$

where,

$$\Pi = (1 - D_1/\beta)(1 - D_2/\beta), \tag{49}$$

$$\Delta = D_1 D_2/\beta^2 - e^{-2(R_1+R_2)}, \tag{50}$$

$$\beta = (\sigma_x^2 \sigma_y^2 - \sigma_{xy}^2)/\sigma_y^2. \tag{51}$$

and $\begin{bmatrix} \sigma_x^2 & \sigma_{xy} \\ \sigma_{xy} & \sigma_y^2 \end{bmatrix}$ is the covariance matrix of the $(X, Y)$. We denote the RHS of (46)-(48) as $D_1^*(R_1)$, $D_2^*(R_2)$ and $D_{12}^*(R_1, R_2, D_1, D_2)$, respectively. The minimum average distortion, $D_{min}^{av}$, when the two descriptions are transmitted over two independent channels with PL probabilities $\mu_1$ and $\mu_2$ is computed as,

$$D_{min}^{av} = \min_{\substack{D_1 > D_1^*(R_1) \\ D_2 > D_2^*(R_2) \\ D_{12} \geq D_{12}^*(R_1,R_2,D_1,D_2)}} \mu_1 \mu_2 \beta + \mu_1 D_1 + \mu_2 D_2 + (1 - \mu_1)(1 - \mu_2) D_{12}. \tag{52}$$

Given that for any $(R_1, R_2, D_1, D_2)$, $D_{12}$ may be set to $D_{12}^*(R_1, R_2, D_1, D_2)$, in (52) this may be replaced. To the best of our knowledge, the rate-distortion bound in the case of symmetric distributed multiple descriptions has not been investigated in the literature so far.

*2) Performance of Asymmetric CDMD*

In Table 1, the rates and the distortions of CDMD encoders, designed with different values of correlation coefficients are investigated and compared. The channels are considered noiseless with a packet loss probability of 0.05. The average distortion when side information is not considered in the source encoder design, but is exploited at the MMSE source decoder, is referred to as $D_{sim,1}^{av}$. When the side information is also not utilized at the decoder, the average distortion is denoted by $D_{sim,2}^{av}$. Also $D_{sim}^{av}$ is the average distortion of a systemS considering SI both at the encoder and decoder design; and $D_{min}^{av}$ is the R-D performance bound obtained using (52).

For simulation results in Table 1 and comparison with the R-D bound when the channels are noiseless, it is assumed that each description is compressed and decompressed independently using an ideal Slepian-Wolf encoder and decoder, respectively. Therefore, the transmission rate pair $(R_1, R_2)$ is equal to $(H(I_1|Y), H(I_2|Y))$.

As evident, when $\rho_{xy}^{real}$ is smaller than 0.95, the performance of the system is only about 2dB away from the rate distortion bound. Comparison of $D_{sim}^{av}$ and $D_{sim,1}^{av}$ demonstrates that a noticeable gain is achieved when the encoder is designed with the proposed method considering the side information. Note that the MD source decoders used for $D_{sim,1}^{av}$ and $D_{sim}^{av}$ are both channel optimized. As expected, the reconstruction distortion of the system which does not utilize the side information at the decoder, $D_{sim,2}^{av}$, is larger than $D_{sim}^{av}$ and $D_{sim,1}^{av}$. The performance improvements in both cases are stronger, when the source and SI are more correlated.

The MDSQ of [16] results in an average distortion of -13.29dB and -16.79dB for $\rho_{xy} = 0$ and $\rho_{xy} = 0.9$, respectively; and the values obtained by the proposed algorithm are equal to -18.65dB and -22.88dB. As observed, optimizing the MD considering the channels and side information improves the performance by about 6dB, and makes it a suitable solution for distributed multiple description coding.

Table 2 presents the side, the central and the average distortions of the proposed asymmetric CDMD over noisy channels. The $\rho_{xy}^{real}$, $\rho_{xy}^{enc}$ and $\rho_{xy}^{dec}$ are set to 0.8. Note that the design objective is to minimize $D^{av}$.

The effect of packet loss probability on the performance of ACDMD is studied in Table 3. The channels are noiseless with packet loss. As evident, the gap of the average distortion from rate distortion bound is increased when the PL probability decreases. This is due to the fact that when PL is small, the signal may be transmitted reliably at lower rates. For example, when there is not any packet loss, the descriptions could be transmitted at a rate of $H(I_1, I_2|Y)$ instead of $H(I_1|Y) + H(I_2|Y)$ without any degradation in performance. For larger packet loss probabilities, the side information plays a stronger role in signal reconstruction and the descriptions are designed more dependent to the SI. This causes a reduction on the transmission rates, $H(I_1|Y)$ and $H(I_2|Y)$.

The effect of quantization of side information on the average distortion of the proposed ACDMD system is studied in Fig. 4. The ACDMD source encoder and decoder are designed based on the exact value of the correlation coefficient between source and SI symbols. The results demonstrate that a 64 level quantizer is sufficient for a wide range of $\rho_{xy}^{real}$.

The sensitivity of the average distortion of an asymmetric CDMD system to the correlation coefficient considered during the CDMD source encoder design, $\rho_{xy}^{enc}$, is studied in Fig. 5. The average distortion is plotted versus the value of $\rho_{xy}^{enc}$. It is obvious from the figure that underestimating the correlation between the source and SI, i.e., $\sigma_{xy}^{enc} < \sigma_{xy}^{real} = \sigma_{xy}^{dec}$, degrades the overall performance of an asymmetric CDMD system less than the case when the correlation is overestimated. This is due to the fact that with overestimating $\sigma_{xy}^{enc}$, the encoder entrusts more than it should to the dependency of the source and SI and hence compresses (lossy) the signal more strongly. As this dependency is in fact weaker at the decoder, the signal is then reconstructed with more than expected distortion.

*B. Results for Symmetric CDMD*

In this section, the performance of the two proposed iterative symmetric CDMD decoders are evaluated. The effect of the number of sources and the amount of channel noise are studied. Also, the performance of the methods presented for selecting the SI sources are assessed and compared. As mentioned, we consider a data gathering wireless sensor network with correlated Gaussian sources in this part.

Fig. 6 demonstrates the performance of the two proposed symmetric CDMD decoders for two networks with 10 and 80 nodes, respectively. The channels are subject to PL with a rate of 0.05. The side information sources are selected using the minimum distortion method. A higher node density leads to the availability of more strongly correlated SI sources for the reconstruction of each of the sources. Hence, as the number of sources in unit area increases, the performance of the ACDMD decoders improves, and therefore, the average reconstruction distortion of the symmetric system decreases. The performance improvement of the soft-SI CDMD decoder over the estimated-SI CDMD decoder is greater, when the source density or the channel noise is stronger. This performance gain can reach up to 3dB.

In Fig. 7, the performance of three methods of minimum physical distance, maximum mutual information and minimum distortion for selecting the side information sources are studied. Both types of the symmetric CDMD decoder are considered. The bit error rate and the packet loss probability of channels are set to 0.005 and 0.05, respectively. The minimum physical distance method for SI selection, although simple, has the worst performance as it does not take into account the channel degradations. Yet, it requires the nodes location information as well. In this case and with possibly poorly selected SI, the two decoding schemes provide comparable performance. As expected, both the maximum mutual information and the minimum distortion methods reduce the average distortion by better SI selection, while the latter provides the best performance. The soft-SI decoder further enhances the performance beyond that due to the estimated-SI decoder in such cases.

### C. Computational Complexity and Memory Usage

The ACDMD decoder is designed for a specific value of correlation between symbols of the source and the SI. Therefore, in a practical system, the continuous value of the correlation coefficient is quantized to $N^{Corr}$ levels in both asymmetric and symmetric CDMD.

For reduced computational complexity, the values of $P(I|Y)$ and $C(I|Y)$ for each $I$, $Y$ and the quantized correlation coefficient are stored at the ACDMD and SI-estimated CDMD decoders. This amounts to, $2N^{Corr}N^{SI}N^{I}$ floating point values, where $N^{I}$ is the number of IA table cells which has been assigned a quantizer index. For the soft-SI CDMD decoder, the values of $\sum_k P(\widetilde{X}_k^s|I^s) P(I^u|\widetilde{X}_k^s)$ and $\sum_k P(\widetilde{X}_k^s|I^s) P(I^u|\widetilde{X}_k^s) C(I^u|\widetilde{X}_k^s)$ in (28) and (32) for each $I^u$, $I^s$ and quantized correlation coefficient are stored at the decoder, i.e., $2N^{Corr}(N^I)^2$ floating point values. Typically $N^I$ is several times smaller than $N^{SI}$, therefore, the soft-SI decoder compared to the estimated-SI decoder needs a storage memory of the same order but smaller in value.

It is seen that the number of operations required for the reconstruction of a symbol of each source at the decoder is $5N^I$, $N^u(5N^I + N^{SI})$ and $N^u(4(N^I)^2 + 4N^I)$ for asymmetric (one source), symmetric ($N^u$ sources) estimated-SI and soft-SI CDMD decoders in each iteration, respectively. Therefore, with typical values of the parameters, the computational complexity of the soft-SI decoder is larger than that of the estimated-SI decoder.

CONCLUSIONS

In this paper, channel optimized distributed multiple description vector quantization schemes are proposed for communications over noisy networks with packet loss. Symbols of multiple correlated sources are separately encoded and jointly reconstructed at a single decoder. First, a channel optimized distributed multiple description encoder is designed using a deterministic annealing approach, which effectively takes into account the dependency between the source and a side information. Simulation results indicate a noticeable gain in comparison to a classic MDVQ or a channel optimized MDVQ, while being only 2dB away from the rate-distortion bound of multiple descriptions with side information at the decoder. The effect of noisy channel and PL is studied; and also it is shown that the ACDMD performance is more sensitive to overestimating the dependency between the source and SI as opposed to its underestimation.

An iterative symmetric CDMD scheme is proposed, whose decoder involves one MMSE ACDMD decoder corresponding to each source and a single side information controller. In this scheme, at the decoder for each source, a SI source is selected. Then, for two setups of estimated-SI and soft-SI decoding, either the reconstructed values or the a posteriori probabilities of the SI source is iteratively used as the side information, respectively. The performance of the symmetric CDMD system is investigated over a data gathering wireless sensor network. The estimated-SI CDMD decoder is less computationally complex in comparison to the soft-SI decoder, however, the latter provides a better performance for communications over noisy channels. Three methods for SI selection is suggested which provide a trade-off of computational complexity and performance for CDMD systems. The proposed schemes suits distributed source coding applications in sensor networks and mono or multi view multimedia compression and communications over noisy channels with packet loss. The particular CDMD-based solutions for such applications and corresponding practical considerations are interesting research challenges in this direction.

## APPENDIX A

Considering the MMSE criterion, the average distortion $D^{av}$ is equal to,

$$D^{av} = E\left[(X - \hat{X})^2\right] = \sum_{all\,Q} P(Q) D_Q. \tag{A.1}$$

As side information is independent from channels, we have,

$$D_Q \stackrel{def}{=} \int f(Y) D_{Y,Q} dY, \tag{A.2}$$

$$D_{Y,Q} \stackrel{def}{=} \sum_{all\,J} \int f(X,J|Y,Q)\left(X - \hat{X}(J|Y,Q)\right)^2 dX, \tag{A.3}$$

where,

$$f(X,J|Y,Q) = \sum_{all\,I} f(X,J|Y,I,Q) P(I|Y,Q) = \sum_{all\,I} f(X|Y,I) P(J|Y,Q) P(I|Y), \tag{A.4}$$

and using the Markov chain of $Y \to X \to \tilde{X} \to I$,

$$f(X|Y,I) = \frac{f(X|Y) P(I|X,Y)}{P(I|Y)} = \frac{f(X|Y)\left[\sum_{all\,k} P(I|X,Y,\tilde{X}_k) P(\tilde{X}_k|X,Y)\right]}{P(I|Y)}$$

$$= \frac{f(X|Y)}{P(I|Y)}\left[\sum_{all\,k} P(I|\tilde{X}_k) P(\tilde{X}_k|X)\right]. \tag{A.5}$$

The term $P(I|Y)$ in (A.4)-(A.5) is written as,

$$P(I|Y) = \sum_{all\,k} P(I,\tilde{X}_k|Y) = \sum_{all\,k} P(I|\tilde{X}_k) P(\tilde{X}_k|Y), \tag{A.6}$$

where,

$$P(\tilde{X}_k|Y) = \int f(X,\tilde{X}_k|Y) dX = \int P(\tilde{X}_k|X) f(X|Y) dX. \tag{A.7}$$

Using (A.3) and (A.4), $D_{Y,Q}$ is written as,

$$D_{Y,Q} = \sum_{all\,J} \sum_{all\,I} P(I|Y) P(J|Y,Q) \int f(X|Y,I)\left(X - \hat{X}(J|Y,Q)\right)^2 dX. \tag{A.8}$$

By defining $C(I|Y)$ as,

$$C(I|Y) \stackrel{def}{=} \int X f(X|Y,I) dX, \tag{A.9}$$

and adding and subtracting it in (A.8), $D_{Y,Q}$ is written as,

$$D_{Y,Q} = D_{Y,Q}^{(1)} + D_{Y,Q}^{(2)}, \tag{A.10}$$

where,

$$D_{Y,Q}^{(1)} \stackrel{\text{def}}{=} \sum_{\text{all } J} \sum_{\text{all } I} P(I|Y)P(J|I,Q) \int f(X|Y,I)(X - C(I|Y))^2 dX \tag{A.11}$$

$$= \sum_{\text{all } I} P(I|Y) \int f(X|Y,I)(X - C(I|Y))^2 dX,$$

$$D_{Y,Q}^{(2)} \stackrel{\text{def}}{=} \sum_{\text{all } J} \sum_{\text{all } I} P(I|Y)P(J|I,Q) \int f(X|Y,I)\left(C(I|Y) - \widehat{X}(J|Y,Q)\right)^2 dX \tag{A.12}$$

$$= \sum_{\text{all } J} \sum_{\text{all } I} P(I|Y)P(J|I,Q)\left(C(I|Y) - \widehat{X}(J|Y,Q)\right)^2.$$

Defining $D^{SE}$ and $D_q^{Ch}$ as below, and using (A.1), the Proposition 1 is proven.

$$D^{SE} \stackrel{\text{def}}{=} \int f(Y) D_{Y,Q}^{(1)} dY, \tag{A.13}$$

$$D_Q^{Ch} \stackrel{\text{def}}{=} \int f(Y) D_{Y,Q}^{(2)} dY. \tag{A.14}$$

## APPENDIX B

The objective in MMSE decoding is to minimize the following distortion,

$$D_{av} = E_{X,J,Y,Q}\left[\left(X - \widehat{X}(J|Y,Q)\right)^2\right]. \tag{B.1}$$

It is easily shown that,

$$\widehat{X}(J|Y,Q) = E_X[X|Y,Q,J] = \int X f(X|Y,Q,J) \, dX. \tag{B.2}$$

Considering the Markov chain $X \to I \to J$, and independency of $X$ and $Q$, $f(X|Y,Q,J)$ is equal to,

$$f(X|Y,Q,J) = \sum_{\text{all } I} f(X|Y,I) P(I|Y,Q,J), \tag{B.3}$$

where $f(X|Y,I)$ is computed using (A.5), and $P(I|Y,Q,J)$ is equal to,

$$P(I|Y,Q,J) = \frac{P(J|Y,I,Q)P(I|Y,Q)}{P(J|Y,Q)} = \frac{P(J|I,Q)P(I|Y)}{P(J|Y,Q)}, \tag{B.4}$$

and,

$$P(J|Y,Q) = \sum_{all\ I} P(J|I,Q)P(I|Y). \tag{B.5}$$

Using (B.2), (B.3) and (A.9),

$$\widehat{X}(J|Q,Y) = \int X \left[\sum_{all\ I} f(X|Y,I)P(I|Y,Q,J)\right] dX = \sum_{all\ I} C(I|Y)P(I|Y,Q,J). \tag{B.6}$$

## APPENDIX C

The probability $P(I^u|Y^u)$ is given by,

$$P(I^u|Y^u) = \sum_{I^s} P(I^u|I^s,Y^u)P(I^s|Y^u). \tag{C.1}$$

The SI in soft-SI CDMD decoder, is provided as $\{P(I^s|Y^s,Q^s,J^s), \forall I^s\}$. Noting the Markov chain, $Y^u \to I^s \to \widetilde{X}^s \to \widetilde{X}^u \to I^u$, the equation (C.1) is then written as,

$$P(I^u|Y^u) = \sum_{I^s} P(I^s|Y^s,Q^s,J^s)P(I^u|I^s) \tag{C.2}$$

$$= \sum_{I^s} P(I^s|Y^s,Q^s,J^s) \sum_k P(I^u|\widetilde{X}^s_k) P(\widetilde{X}^s_k|I^s).$$

The proof of other equations in Proposition 3 is straightforward.

## APPENDIX D

The equation (31) could be obtained similar to equation (17), and using (13), $C(I^u|Y^u)$ is written as,

$$C(I^u|Y^u) = \int X^u f(X^u|Y^u,I^u) dX^u. \tag{D.1}$$

Similar to (A.5), we have,

$$f(X^u|Y^u,I^u)P(I^u|Y^u) = \left[\sum_l P(I^u|\widetilde{X}^u_l) P(\widetilde{X}^u_l|X^u)\right] f(X^u|Y^u). \tag{D.2}$$

In the proposed CDMD decoder with soft-SI $Y^u$, $P(I^s|Y^u)$ is given by the a posteriori probability $P(I^s|Y^s,Q^s,J^s)$. We have

$$f(X^u|Y^u) = \sum_{I^s} f(X^u,I^s|Y^u) = \sum_{I^s} P(I^s|Y^u) f(X^u|I^s) \tag{D.3}$$

$$= \sum_{I^s} P(I^s|Y^s,Q^s,J^s) f(X^u|I^s)$$

$$= \sum_{I^s} P(I^s|Y^s, Q^s, J^s) \sum_k P(\widetilde{X}_k^s|I^s) f(X^u|\widetilde{X}_k^s).$$

Also it is straightforward to show that,

$$f(X^u|\widetilde{X}_k^s, I^u) = \frac{f(X^u|\widetilde{X}_k^s)}{P(I^u|\widetilde{X}_k^s)} \left[ \sum_l P(I^u|\widetilde{X}_l^u) P(\widetilde{X}_l^u|X^u) \right]. \tag{D.4}$$

Using (D.3) and (D.4), the equation (D.2) is written as,

$$P(I^u|Y^u) f(X^u|Y^u, I^u) = \sum_{I^s} P(I^s|Y^s, Q^s, J^s) \sum_k P(\widetilde{X}_k^s|I^s) P(I^u|\widetilde{X}_k^s) f(X^u|\widetilde{X}_k^s, I^u). \tag{D.5}$$

Using the definition of $C(I^u|\widetilde{X}_k^s)$ in (33) and equations (D.2) and (D.5), the equation (32) is obtained. The proof of the equation (34) in Proposition 4 is straightforward.

## APPENDIX E

The mutual information in (36), is written as,

$$I(J^u, J^t|Q^u, Q^t) = H(J^u|Q^u) + H(J^t|Q^t) - H(J^u, J^t|Q^u, Q^t), \tag{E.1}$$

where,

$$H(J^u|Q^u) = -\sum_{J^u} P(J^u|Q^u) \log P(J^u|Q^u), \tag{E.2}$$

$$H(J^u, J^t|Q^u, Q^t) = -\sum_{J^u} \sum_{J^t} P(J^u, J^t|Q^u, Q^t) \log P(J^u, J^t|Q^u, Q^t). \tag{E.3}$$

Considering the Markov chain of $J^u \to I^u \to \widetilde{X}^u \to \widetilde{X}^t \to I^t \to J^t$, the probabilities $P(J^u|Q^u)$ and $P(J^u, J^t|Q^u, Q^t)$ are written as,

$$P(J^u|Q^u) = \sum_{I^u} P(J^u|I^u, Q^u) \sum_l P(I^u|\widetilde{X}_l^u) P(\widetilde{X}_l^u), \tag{E.4}$$

$$P(J^u, J^t|Q^u, Q^t) = \sum_{I^u} \sum_{I^t} P(J^u|I^u, Q^u) P(J^t|I^t, Q^t) \sum_l \sum_k P(I^u|\widetilde{X}_l^u) P(I^t|\widetilde{X}_k^t) P(\widetilde{X}_l^u, \widetilde{X}_k^t). \tag{E.5}$$

The equations (E.4) and (E.5) include averaging on the noise of the description channels. Therefore, the PL status of the channels and the statistics of channels noises are taken into account to compute the desired mutual information.

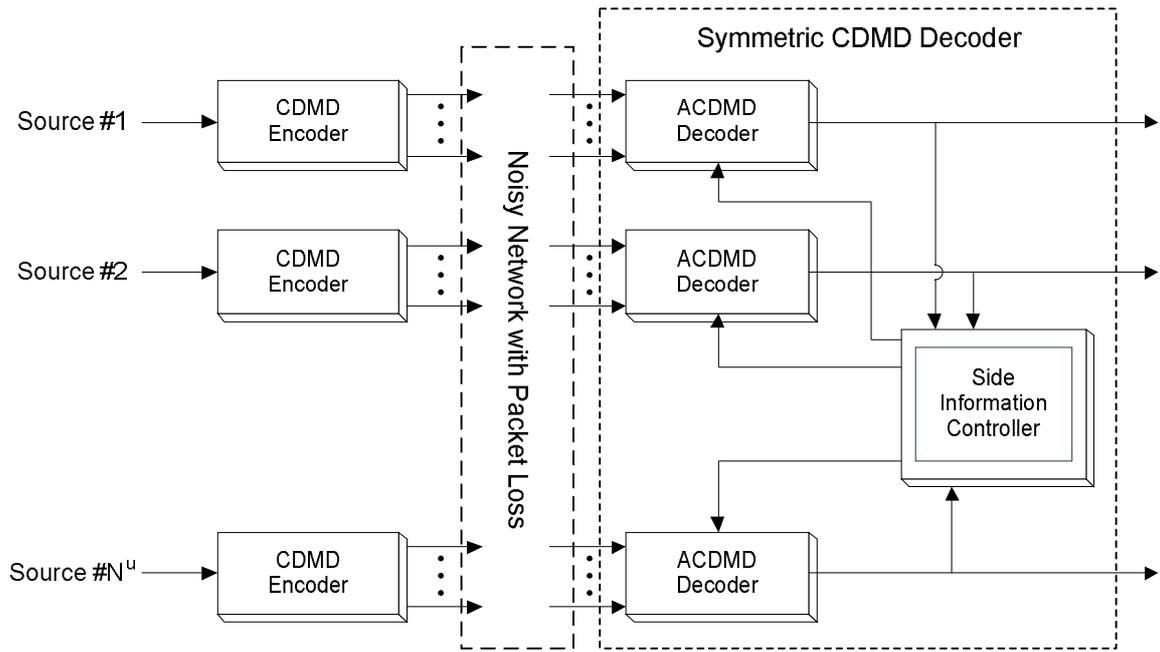

Fig. 1. System model of a symmetric CDMD system

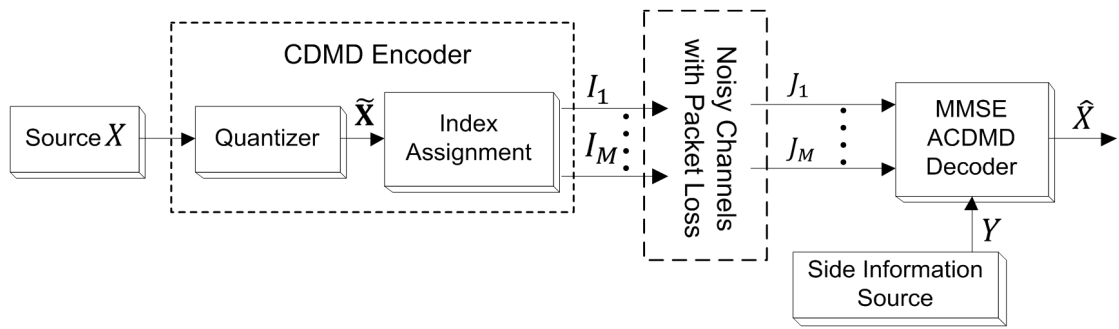

Fig. 2. System model of an asymmetric CDMD

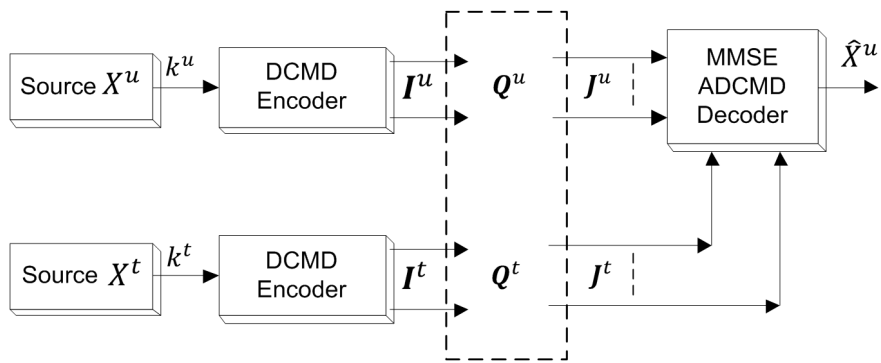

Fig. 3. Asymmetric CDMD system with noisy partial side information

TABLE 1
EFFECT OF CDMD ENCODER DESIGN ON PERFORMANCE OF ACDMD SYSTEM
($\mu_m = 0.05$, noiseless channel, distortion in [dB])

| $\rho_{xy}$ | $R_1$ | $R_2$ | $D_{sim}^{av}$ | $D_{min}^{av}$ | $D_{sim}^{av} - D_{min}^{av}$ | $D_{sim,1}^{av}$ | $D_{sim}^{av}$-$D_{sim,1}^{av}$ | $D_{sim,2}^{av}$ | $D_{sim}^{av}$-$D_{sim,2}^{av}$ |
|---|---|---|---|---|---|---|---|---|---|
| 0 | 2.80 | 2.81 | -18.654 | -20.509 | 1.855 | -18.654 | 0 | -18.653 | 0 |
| 0.2 | 2.78 | 2.78 | -18.696 | -20.565 | 1.869 | -18.694 | 0.002 | -18.653 | 0.042 |
| 0.4 | 2.71 | 2.69 | -18.827 | -20.784 | 1.957 | -18.823 | 0.005 | -18.653 | 0.173 |
| 0.6 | 2.54 | 2.53 | -19.160 | -21.188 | 2.028 | -19.091 | 0.067 | -18.653 | 0.507 |
| 0.8 | 2.32 | 2.32 | -20.619 | -22.608 | 1.989 | -19.714 | 0.906 | -18.653 | 1.965 |
| 0.9 | 2.25 | 2.22 | -22.882 | -24.935 | 2.053 | -20.4204 | 2.462 | -18.653 | 4.228 |
| 0.95 | 2.20 | 2.22 | -25.576 | -27.689 | 2.113 | -21.2014 | 4.375 | -18.653 | 6.922 |
| 0.99 | 2.16 | 2.16 | -30.894 | -34.327 | 3.433 | -23.687 | 7.208 | -18.653 | 12.241 |

TABLE 2
PERFORMANCE OF ASYMMETRIC CDMD OVER BSC CHANNELS
($R_1 = R_2 = 3$, $\mu_m = 0.05$, $\rho_{xy} = 0.8$, distortion in [dB])

| $P_1 = P_2$ | $(D_1 + D_2)/2$ | $D_{12}$ | $D^{av}$ |
|---|---|---|---|
| 0.1 | -8.102 | -10.946 | -10.546 |
| 0.01 | -11.631 | -19.244 | -17.407 |
| 0.001 | -12.877 | -22.858 | -19.799 |
| 0.0001 | -13.084 | -24.285 | -20.489 |
| 1.00E-05 | -13.101 | -24.592 | -20.612 |
| 0 | -13.109 | -24.601 | -20.619 |

TABLE 3
EFFECT OF PACKET LOSS ON THE PERFORMANCE OF ASYMMETRIC CDMD
(Noiseless channel, $\rho_{xy} = 0.8$, distortion in [dB])

| $\mu_1 = \mu_2$ | $R_1$ | $R_2$ | $D_{sim}^{av}$ | $D_{min}^{av}$ | $D_{sim}^{av} - D_{min}^{av}$ |
|---|---|---|---|---|---|
| 0.3 | 2.265 | 2.269 | -13.289 | -13.758 | 0.469 |
| 0.2 | 2.28 | 2.259 | -15.605 | -16.365 | 0.76 |
| 0.1 | 2.276 | 2.271 | -18.496 | -19.896 | 1.400 |
| 0.05 | 2.321 | 2.319 | -20.619 | -22.608 | 1.989 |
| 0.02 | 2.389 | 2.498 | -22.497 | -25.751 | 3.254 |
| 0.01 | 2.459 | 2.53 | -24.206 | -27.622 | 3.416 |
| 0.005 | 2.635 | 2.546 | -25.149 | -29.676 | 4.527 |

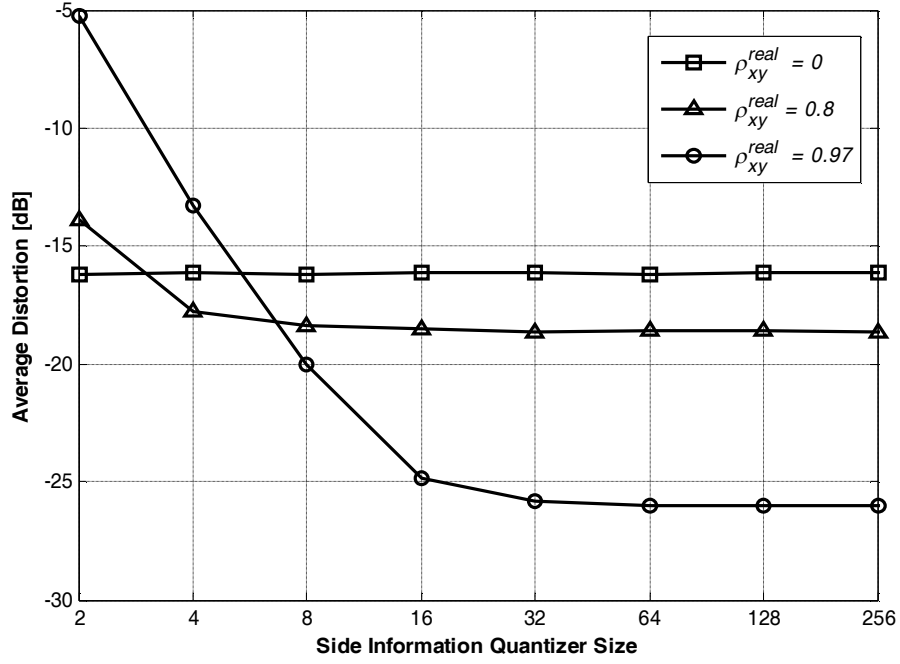

Fig. 4. Sensitivity of CDMD to side information quantizer size
(BSC channels, $P_m = 0.005$, $\mu_m = 0.05$, distortion in [dB])

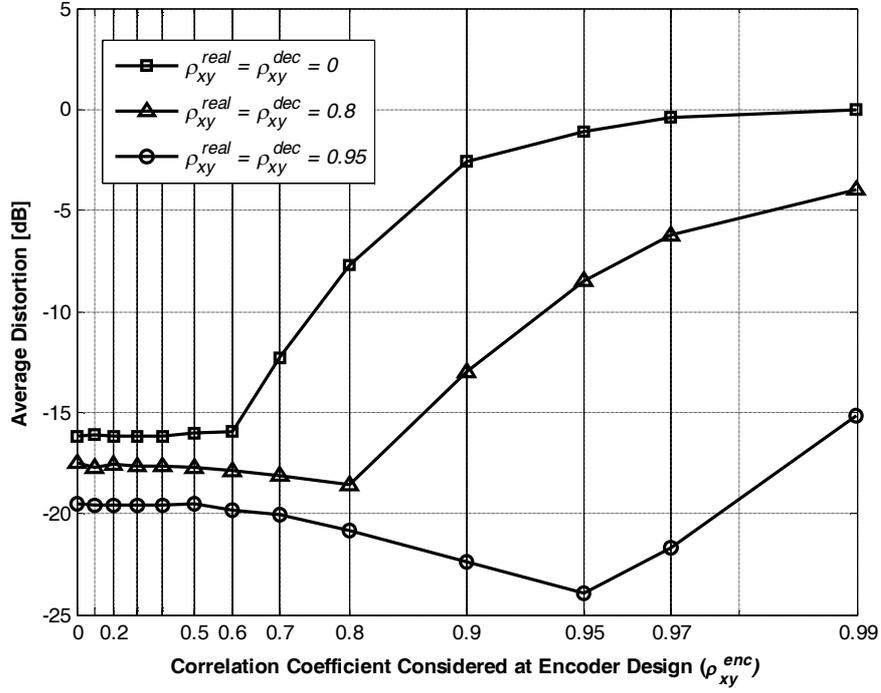

Fig.5. Sensitivity of performance of ACDMD to mismatch between $\rho_{xy}^{enc}$ and $\rho_{xy}^{real}$
(BSC channels, $P_m = 0.005$, $\mu_m = 0.05$, distortion in [dB])

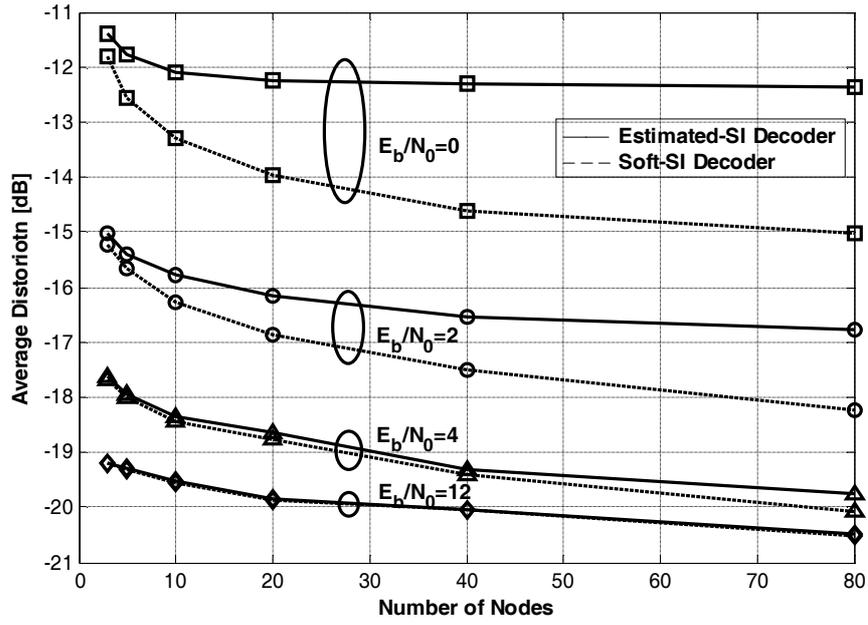

Fig. 6. Performance of symmetric CDMD decoder with different number of nodes

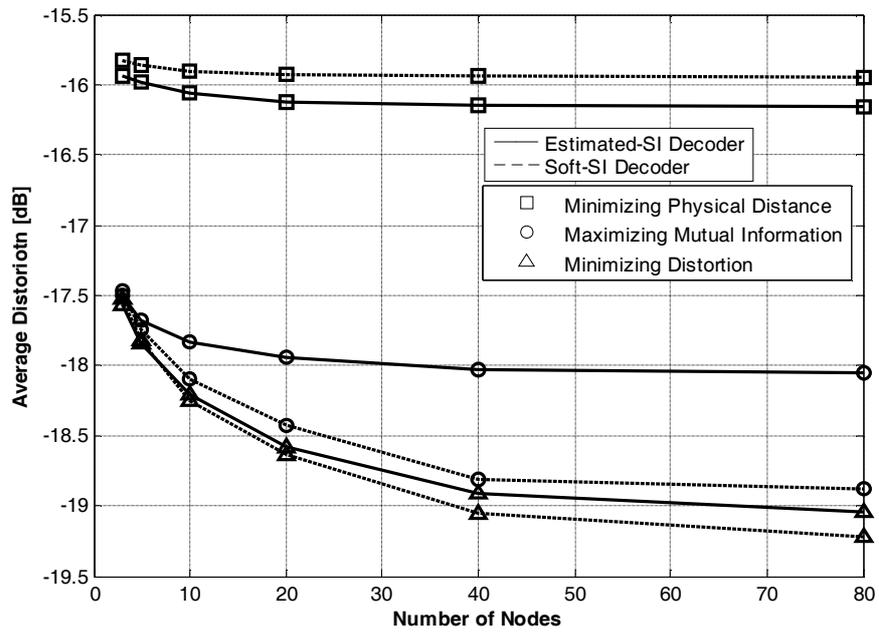

Fig. 7. Comparison of methods of selecting side information sources at symmetric CDMD decoder

(BSC channels, $P_m = 0.005$, $\mu_m = 0.05$, distortion in [dB])